\documentclass[aps,pre,superscriptaddress,twocolumn]{revtex4-1}

\usepackage{graphicx}
\usepackage[utf8]{inputenc}
\usepackage[english]{babel}
\usepackage{amsmath,graphicx,enumerate}
\usepackage{epsfig}
\usepackage{hyperref}
\usepackage{amssymb}
\usepackage{color}
\usepackage{float} 

\DeclareRobustCommand{\vect}[1]{
 \ifcat#1\relax
 \boldsymbol{#1}
 \else
 \mathbf{#1}
 \fi}
 
 \DeclareRobustCommand{\matr}[1]{
 \mathsf{#1}
 }

\begin{document}
\title{Response and flux of information in extended non-equilibrium dynamics
}
\author{Camilla Sarra}
\affiliation{Department of Physics, Princeton University, Princeton, New Jersey 08544, USA}
\author{Marco Baldovin}
\affiliation{Dipartimento di Fisica, Universit\`a di Roma Sapienza, P.le Aldo Moro 5, 00185, Rome, Italy}
\author{Angelo Vulpiani}
\affiliation{Dipartimento di Fisica, Universit\`a di Roma Sapienza, P.le Aldo Moro 5, 00185, Rome, Italy}

\begin{abstract}
It is well known that entropy production is a proxy to the detection of 
non-equilibrium, i.e. of the absence of detailed balance; however, due to the 
global character of this quantity, its knowledge does not allow to identify 
spatial currents or fluxes of information among specific elements of the system 
under study. In this respect, much more insight can be gained by 
studying transfer entropy and response, which allow quantifying the relative 
influence of parts of the system and the asymmetry of the fluxes.
In order to understand the relation between the above-mentioned quantities, we 
investigate spatially asymmetric extended systems. First, we consider a 
simplified linear stochastic model, which can be studied analytically; then, we 
include nonlinear terms in the dynamics. Extensive numerical investigation 
shows the relation between entropy production and the 
above-introduced degrees of asymmetry. Finally, we apply our approach to the 
highly nontrivial dynamics generated by the Lorenz '96 model for Earth oceanic 
circulation. 
\end{abstract}

\maketitle

\section{Introduction}
\label{sec:introduction}
From a physical point of view, one of the most relevant aspects of 
out-of-equilibrium systems is the presence of currents induced by external 
constraints or fields~\cite{degroot84,livi2017nonequilibrium}. These currents have the dual effect of breaking 
the time-reversal symmetry of the system and of producing spatial inhomogeneity 
between the variables. For instance, an electric field acting on charged 
particles induces a mean current, in the same way as imposing different 
temperatures at the end of a bar induces a heat flux ruled by Fourier's law. 

In mathematical terms, a system is out of equilibrium if the detailed balance 
does not hold, or, equivalently, if the entropy production $\Sigma$ is 
positive~\cite{seifert2012stochastic}. It is well known that in the very general context of Markov 
processes (e.g. Langevin equations and master equations) an explicit expression 
for $\Sigma$ can be introduced by comparing the probability of a long trajectory 
to that of its time reversal~\cite{gallavotti-cohen}.
One may ask whether and  in which sense the lack of time-reversal symmetry 
measured by $\Sigma$ is related to the presence of currents of physical 
observables, i.e. to the breaking of the spatial symmetry of the
dynamics~\cite{loos2020thermodynamic}. Let 
us stress, for instance, that even low dimensional systems, as $2D$ linear 
Langevin equations, can have a non-zero entropy production and be consistently 
classified as out-of-equilibrium; it is quite obvious, however, that it is 
misleading to use concepts as currents and spatial symmetry breaking in these 
cases, because of the absence of a spatial structure~\cite{crisanti2012nonequilibrium}.

In this paper, we investigate high-dimensional extended systems, analyzing both 
the temporal and spatial aspects of out-of-equilibrium states. In particular, to 
characterize the breaking of the spatial symmetry we adopt two approaches, 
namely the study of transfer entropy (TE), borrowed from information theory, and 
the analysis of response functions (RF). In the former case, the idea is to 
quantify the amount of information exchanged between two variables $x_i$ and 
$x_j$ of the system, i.e., roughly speaking, the improvement in our ability to 
predict $x_j$ once $x_i$ is known. The latter approach consists instead of 
measuring how the perturbation of a variable $x_i$ influences on average the 
behavior of $x_j$. 

The paper is organized as follows. In Section~\ref{sec:aysmmetry} we discuss in 
some detail the observables we will study in the rest of the paper, namely 
$\Sigma$, TE and RF. In Section~\ref{sec:ring} we introduce a system of $N$ 
linear ``oscillators'' with asymmetric interactions, ruled by a Markov dynamics. This 
exactly solvable system allows us to study the relation between non-equilibrium 
and symmetry breaking. The case with some nonlinear terms in the interactions is 
also considered. Section~\ref{sec:lorenz} is devoted to the analysis of the 
so-called Lorenz 96 model, i.e. a simplified description of the dynamics of 
representative atmospheric observables. Despite the apparent simplicity, such a 
system contains the main difficulties of turbulence and only numerical 
computations are possible. By varying the parameters of the model (dissipation and 
external forcing) we give a qualitative scenario of the crossover from 
equilibrium to non-equilibrium situations. 
Our conclusions are drawn in Section~\ref{sec:conclusions}.

\section{Characterizing non-equilibrium and spatial asymmetry}
\label{sec:aysmmetry}
To understand to what extent a certain system is out of equilibrium, we need to quantify the asymmetry of its dynamics under time-reversal transformations, i.e. its irreversibility, due to the breaking of the detailed balance condition.
Consider the direct trajectory $\{x(s)\}_{0<s<t}$ of a time-dependent variable $x(s)$ from time $s=0$ to time $s=t$; let us denote by $P(\{x(s)\}_{0<s<t})$ the probability of $\{x(s)\}_{0<s<t}$ and by $\{\mathcal I x(s)\}_{0<s<t}$ the reverse trajectory, which collects the same states as the direct trajectory but in the reversed order. 
As it is well known, the functional~\cite{lebowitz1999gallavotti, malek2017fluctuation} 
\begin{equation}
    \label{Lebowitz-Spohn} \Delta S(t) = \ln {\frac{ P (\{x(s)\}_{0<s<t})}{ P (\{\mathcal{I} x(s)\}_{0<s<t})} }\,,
\end{equation}
can be proved to be the cumulative entropy production along the trajectory.

In some particular systems, the quantity in Eq.~\eqref{Lebowitz-Spohn} can be 
explicitly written down for the single trajectory by evaluating its probability, 
according to the path-integral procedure introduced by Onsager and 
Machlup~\cite{onsager1953fluctuations}. To be more specific, let us focus on the 
kind of systems we will deal with in the following Sections, which are Markovian 
stochastic systems described by an $N$-dimensional Langevin equation, such as 
\begin{equation}
\label{model_system} \dot{\mathbf{x}} = \mathbf{F}(\mathbf{x}) 
+ \matr{ \matr{B}} \vect{\eta}\,,
\end{equation}
where $\mathbf{F}: \mathbb{R}^N \mapsto \mathbb{R}^N$ is a smooth function, 
$\matr{B}$ is the diagonal matrix with elements $B_{ij}=\delta_{ij}\sqrt{2T_i}$, 
and $\vect{\eta}$ represents a white delta-correlated noise, such that 
$\left\langle \eta_i(t) \eta_j(s)\right\rangle= \delta_{ij} \delta(t-s)$. If we 
consider the initial and final state as stationary states, it is possible to 
show~\cite{lebowitz1999gallavotti} that the mean entropy production rate reads 
\begin{equation}
\Sigma= \sum_k \frac{\langle F_k\left(\mathbf{x}\right)\dot{x}_k\rangle}{T_k}\,.
\end{equation}
In the systems in which we can consider Eq.~\eqref{model_system} as the motion 
of a particle in a medium, we can get a more physical intuition of the above 
result~\cite{seifert2005entropy, harris2007fluctuation}.

This paper aims to explore the relation between non-equilibrium and spatial 
asymmetry in paradigmatic models. While a handy tool to measure the lack of 
equilibrium is naturally provided by entropy production, finding a suitable way 
to characterize spatial asymmetry is a much less obvious task. 

The first approach that we shall follow is based on the assumption that spatial asymmetry in the dynamics is realized when some kind of ``effective interaction'' between two variables is no longer symmetric, i.e. when a 
perturbation of variable $x_i$ induces on $x_j$ an average effect which is different from that exerted on $x_i$ when perturbing $x_j$. 
The physical observable which quantifies this average effect is the response function, which plays an important role in the study of both equilibrium and out-of-equilibrium systems.
Given an $N$-dimensional system $\vect{x}(t)$, the response of variable $x_i$ to an instantaneous perturbation $\delta x_j(0)$ of $x_j$ at time $0$ is defined as
\begin{equation}
\matr{R}_{ij}(t)=\frac{\overline{\Delta x_i(t)}}{\delta x_j(0)}
\end{equation}
where $\Delta x_i(t)$ is the difference between the perturbed and the unperturbed trajectory, and $\overline{\,\cdot\,}$ is an average over a large number of realizations of the experiment.
In what follows, we shall analyze the breaking of spatial symmetry by measuring the asymmetry of response functions integrated over time.

The usefulness of response is mainly due to a number of fluctuation-dissipation theorems establishing analytical relations between response functions and correlations of suitable observables when the perturbation is small~\cite{kubo1966fluctuation, marconi2008fluctuation}.
 Under very general hypotheses, a fluctuation-dissipation theorem valid also for out-of-equilibrium systems can be formulated~\cite{sarracino2019fluctuation}. 
Denoting by $\rho(\mathbf{x})$ the stationary probability density function (pdf), such generalized fluctuation-dissipation relation reads
\begin{equation}
\label{eq:fdr}
 \matr{R}_{ij}(t)=-\left\langle x_i(t) \frac{\partial }{\partial x_j} \log[\rho(\mathbf{x}(0))]\right\rangle\,.
\end{equation}

When the stationary solution is a Gaussian, as it happens for linear models with 
additive Gaussian noise, the theorem assumes a particularly simple form and 
response functions can be written as a sum of correlations, according to 
\begin{equation}
\matr{R}_{ij}(t) = \sum_k\matr{C}_{ik}(t) [ \matr{\sigma}^{-1}]_{kj}\,,
\end{equation}
where  $ \matr{C}=\left\langle \vect{x}(t) \vect{x}^{T}(0)\right \rangle$ is the
correlation matrix and $ \matr{\sigma} = \matr{C}(t=0)$ is the covariance matrix. 
The above formula can be thus written in matrix form as the linear regression relation
\begin{equation}\label{intro-linear_regression}
 \matr{R}(t) = \matr{C}(t) \matr{\sigma}^{-1}.
\end{equation}

A different approach to characterize spatial asymmetry in the dynamics may be 
focused on the currents of probability. Indeed, currents break the spatial 
equivalence between the different parts of a system and introduce a preferred 
direction, which is the one corresponding to the flux. A quantity that can be 
used to study this effect is the transfer entropy (TE), which is used in 
information theory to measure the information exchanged between time-dependent 
variables~\cite{schreiber2000measuring, san2007rigorous,hlavavckova2007causality,barnett2009granger}. 

Consider two random variables $x$ and $y$, and let us indicate with $P_{xy} 
(x,y)$ their joint probability and with $P_x(x)$ and $P_y(y)$ the marginal 
distributions.
The Shannon entropy of variable $x$ is defined as
\begin{equation}\label{EQ:shannon_entropy}
 H(x) \equiv -\int dx P_x(x) \ln P_x(x)\,,
\end{equation}
while the joint Shannon entropy of $x$ and $y$ reads
\begin{equation}\label{EQ:joint_entropy}
 H(x,y) \equiv -\int dx dy P_{x y} (x,y) \ln P_{x y}(x,y).
\end{equation}
It follows that if $x$ and $y$ are independent, then $H(x,y) = H(x)+H(y)$.
The conditional entropy of $x$ given $y$ is consistently defined as 
\begin{align}\label{EQ:conditional_entropy}
    H(x|y) &\equiv H(x,y) - H(y)
\end{align}
and can be interpreted as the uncertainty about the value of $x$ once $y$ is known. 
From what we said, for independent variables $H(x|y)=H(x)$.
The amount of information shared between $x$ and $y$ is quantified by the mutual information 
\begin{equation}\label{EQ:mutual_info}
 I(x,y) \equiv H(x) + H(y) - H(x,y)\,.
\end{equation}
$I(x,y)$ is always positive, symmetric under the exchange of its two 
arguments and it is zero if and only if the variables are independent.
Transfer entropy is a particular time-asymmetric conditional mutual information;
for Markovian systems in discrete time it is defined as
\begin{equation}\label{def:transfer_entropy}
\begin{split}
 \mathcal{T}_{y \to x}(t) &\equiv I\left(x_t, y_{t-1} |x_{t-1}\right) =\\ 
 &= H(x_t|x_{t-1}) - H(x_t|x_{t-1},y_{t-1})\,.
\end{split}
\end{equation}
For stationary systems, which are the only ones we will deal with in the following Sections, the transfer entropy does not depend on time, so $\mathcal{T}_{y\to x}(t) = \mathcal{T}_{y\to x}$.

We point out that some attention is needed with continuous-time systems \cite{bossomaier2016introduction}.
The meaningful quantity in this case is the information transfer rate, i.e. the amount of information transferred per unite time:
\begin{equation}\label{continuous_time_TE}
 \mathcal{T}_{y \to x}(t) = \lim_{\delta t\to 0} \frac{I(x(t),y(t-\delta t)|x(t-\delta t))}{\delta t}.
\end{equation}
For the systems we want to focus on in the following Sections, it can be explicitly seen that this limit is well-defined and, if $\delta t$ is small enough, the quantity on the r.h.s. of Eq.~\eqref{continuous_time_TE} is almost independent of it.

In order to compute TE, from a practical point of view, one needs to evaluate conditional entropies. This is feasible by means of explicit calculations when the stationary distribution is known: a relevant example, which will be considered in the following Section, is that of Gaussian distributions, detailed in Appendix~\ref{sec:appendix_te}.
In most cases, however, the stationary distribution is not known and hence we do not have an explicit form for the conditional entropies. In these cases one has to rely on numerical estimations of entropy from data, which might not be a trivial task in the general case.

\section{Asymmetric ring models}
\label{sec:ring}

In this Section we focus on ring models, i.e. extended systems of $N$ variables $\{x_i\}_{i=1,...N}$ whose dynamics is ruled by the stochastic process 
\begin{equation}
\label{eq:genmodel}
 \dot{x_i}=F(x_i, x_{i-1},x_{i+1})+ \sqrt{2 T_i} \eta_i\quad \forall\, i
\end{equation} 
with
\begin{equation}
\label{eq:drift}
 F(x,y,z)=-f(x)-f_L(x-y)+f_R(z-x)
\end{equation} 
where $f$, $f_L$ and $f_R$ are smooth real-valued functions, $\{T_i\}_{i=1,...N}$ are positive constants and $\{\eta_i\}_{i=1,...N}$ represent $N$ independent Gaussian noises with zero mean satisfying $\langle \eta_i(t) \eta_j(s)\rangle=\delta(t-s)\delta_{ij}$ . Periodic boundary conditions are assumed, i.e. $x_{N+1}\equiv x_{1}$ and $x_{0}\equiv x_{N}$.

We shall first consider an exactly solvable case, i.e model~\eqref{eq:genmodel} with linear $f$, $f_L$ and $f_R$. This can be seen as the overdamped limit of a system of coupled linear oscillators with asymmetric interactions (see Ref.~\cite{ishiwata20} for a detailed analytical discussion of the underdamped case).
After that, the effect of nonlinearities will be taken into account.

\subsection{Linear cases}
\label{sec:linear}

If we assume that the drift terms appearing in Eq.~\eqref{eq:drift} are linear, i.e.
$$ f(r)=\gamma r\,,\quad f_L(r)=\alpha r\,,\quad f_R(r)=\beta r\,,\quad$$ 
for some constants $\alpha$, $\beta$ and $\gamma$, the model we are interested in can be written as
\begin{equation}
 \dot{\vect{x}}=-\matr{A}\vect{x}+\matr{B}\vect{\eta}
\end{equation} 
where $\matr{A}$ is the Toepliz matrix whose elements are given by
\begin{equation}
\matr{A}_{ij}=(\alpha+\beta+\gamma) \delta_{i,j}-\alpha \delta_{i,j-1}-\beta \delta_{i,j+1}
\end{equation} 
(periodic boundary conditions are assumed) and
\begin{equation}
 \matr{B}_{ij}=\sqrt{2 T_i}\delta_{ij}\,.
\end{equation} 
It is useful to define the symmetric and antisymmetric part of the interactions with ``neighbor'' variables as $$ k=\frac{\beta+\alpha}{2}\quad \text{and} \quad \varepsilon=\frac{\beta-\alpha}{2k}\,$$ 
respectively.

\begin{figure}
 \centering
 \includegraphics[width=.8\linewidth]{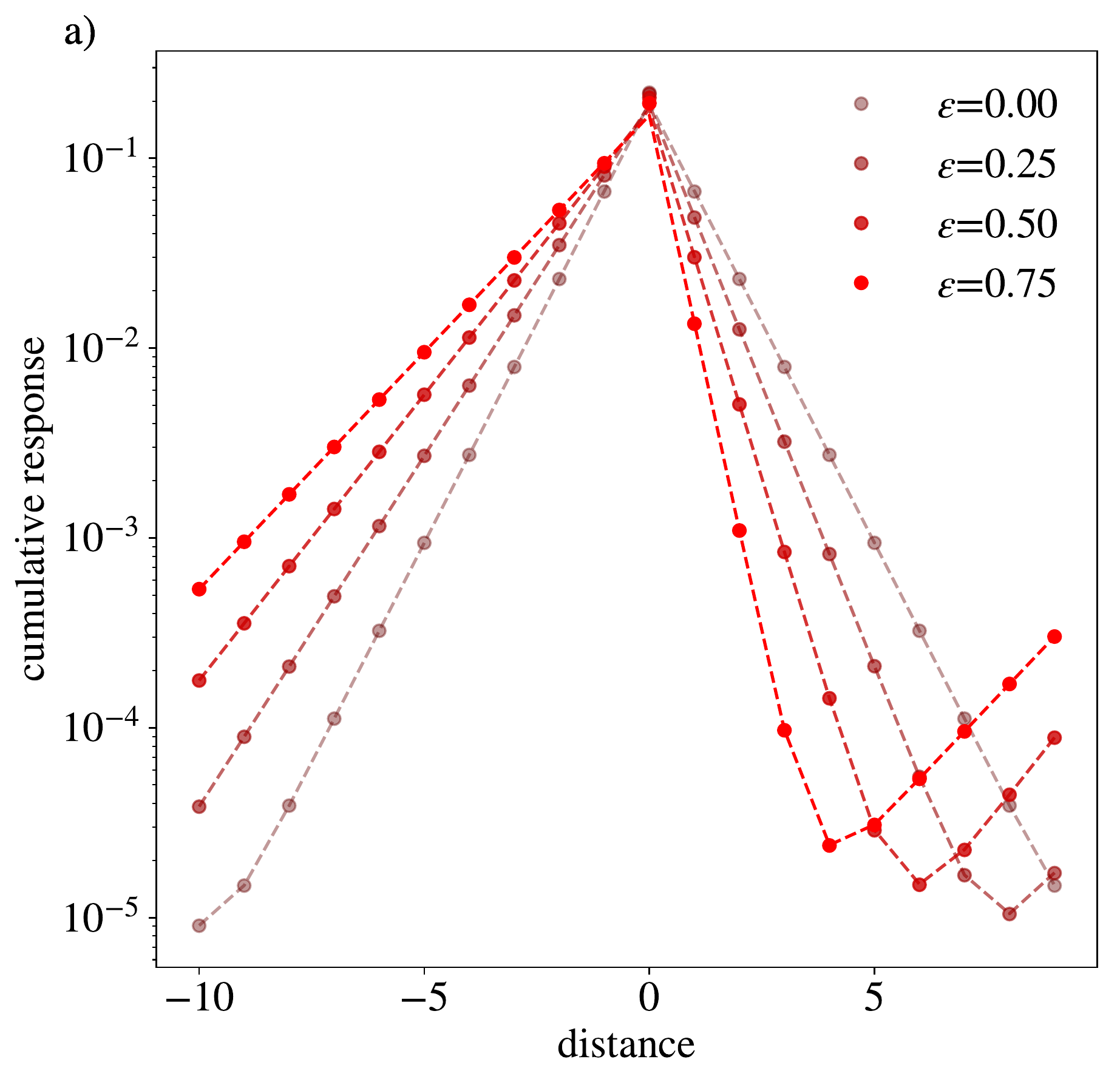}
 \includegraphics[width=.8\linewidth]{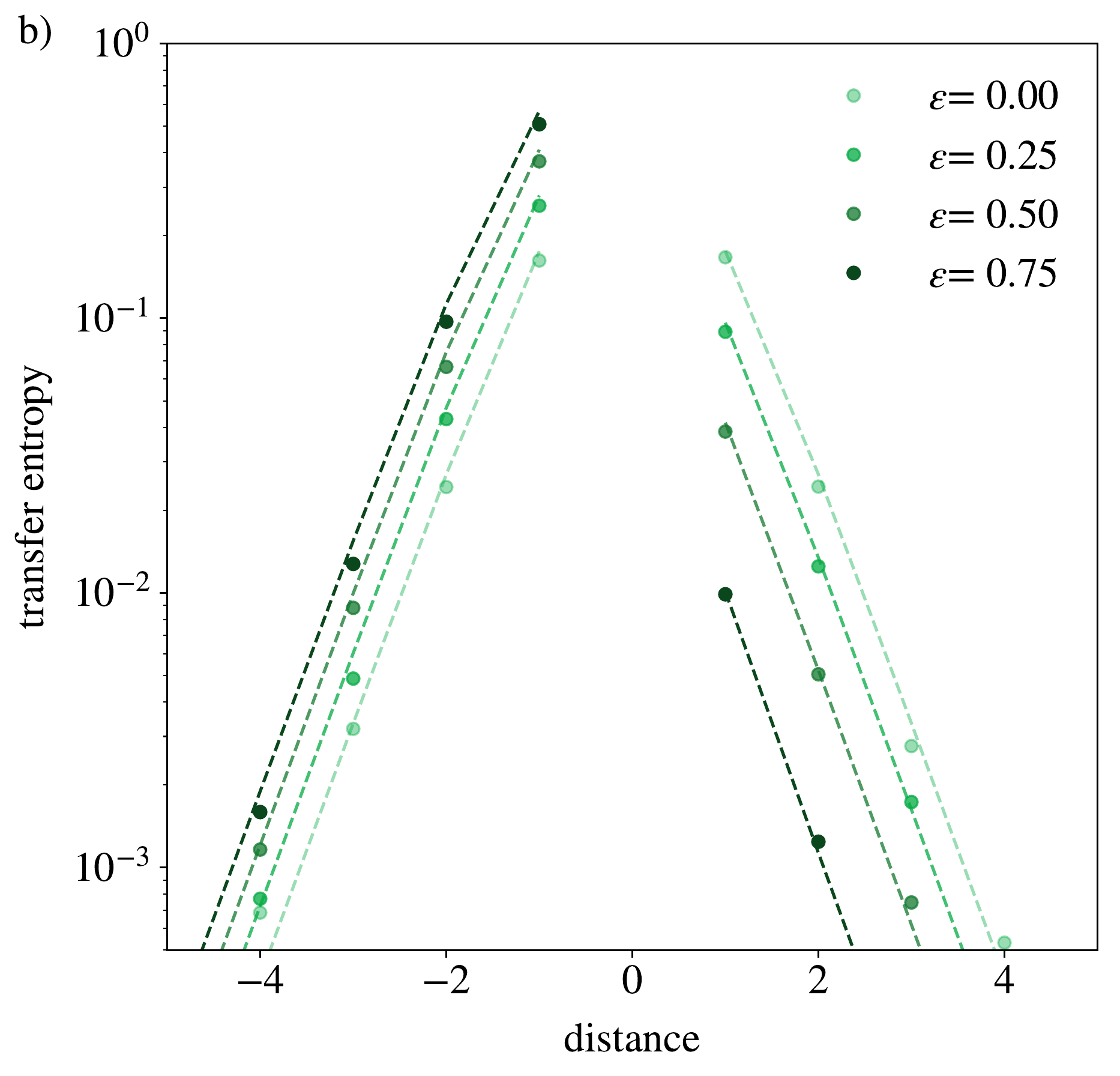}
 \caption{Spatial asymmetry in the linear case with uniform noise. 
 Cumulative response $\tilde{\matr{R}}_{ji}$ [panel (a)] and transfer entropy $\mathcal{T}_{i\to j}$ [panel (b)] are plotted as functions of the oriented distance $(j-i)$, for different values of the parameter $\varepsilon$ ruling the asymmetry of the interactions. Circles represent numerical simulations, dashed lines are theoretical predictions. 
 Parameters: $k=2$, $\gamma = 2.5$, $T=0.5$,  $N=20$.
 In panel (a): time-step $\Delta t = 10^{-3}$, integration time $\tau = 5$,  perturbation $\delta x=0.1$, average over $M=1000$ realizations. 
 In panel (b): $\Delta t = 10^{-2}$, $M=5 \cdot 10^4$.
 \label{fig:linear}}
\end{figure}

For this class of linear models, we can exploit exact relations for the 
computation of $\Sigma$ and RFs, as discussed in the previous section, and also 
of TEs, as detailed in Appendix~\ref{sec:appendix_te}. Similar studies on systems
with asymmetric linear interactions have been performed, for instance, in Ref~\cite{ishiwata20}.
First we focus on the 
case in which all noises have equal amplitudes, i.e.
\begin{equation}   
\label{eq:linear_1}
T_i=T\quad \forall\, i\,.
\end{equation}
In Figure~\ref{fig:linear}(a) we fix a variable $x_i$ and we compute the ``cumulative'' response
\begin{equation}
 \tilde{\matr{R}}_{ji}=\int_{0}^{\infty} dt \, \matr{R}_{ji}(t)\,,
\end{equation} 
inspired by the Green-Kubo formulas, for several values of the ``oriented distance'' $(j-i)$ between the considered 
variables. This can be achieved numerically by comparing the evolution of an 
``unperturbed'' trajectory with that obtained by implementing the instantaneous 
transformation \begin{equation}
 x_i(0) \to x_i(0) +\delta x_i(0)\,.
\end{equation}
The function $\matr{R}_{ji}(t)$ obtained in this way is integrated 
in the time interval $[0,\tau]$, where $\tau$ is much larger than
the typical characteristic times of the dynamics (see figure caption for details).
The results of such numerical simulations are compared with the analytical 
relation~\eqref{eq:fdr}, which yields of course
\begin{equation}\label{lin-regr-cumu}
 \tilde{\matr{R}} \simeq \tilde{\matr{C}} \sigma^{-1}\,.
\end{equation}

The experiment is repeated for several values of 
$\varepsilon$. Similarly, in Fig.~\ref{fig:linear}(b) the TE is numerically 
computed and compared with the analytical prediction for the linear case. We 
point out that, as explained in Appendix~\ref{sec:appendix_te}, in this case 
transfer entropies can be written in terms of covariances and correlations. At 
variance with response functions, TE is trivially zero when $j=i$ (see 
Eq.~\eqref{def:transfer_entropy}).

Both observables show, as expected, a symmetric behavior when $\varepsilon$, the 
anti-symmetric part of the interaction, is zero; for positive values of 
$\varepsilon$ the RF and the TE from $x_i$ to $x_{i+n}$ with $0<n<N/2$ are 
larger than those from $x_i$ to $x_{i-n}$. However, it should be noticed that as 
far as RFs are concerned, varying $\varepsilon$ amounts to a change in the 
decay length; the effect on TE is instead only limited to the presence of an 
overall rescaling factor, leaving untouched the slope of the RF in the semi-log
plot in Fig.~\ref{fig:linear}(b).

\begin{figure}
 \centering
 \includegraphics[width=.8\linewidth]{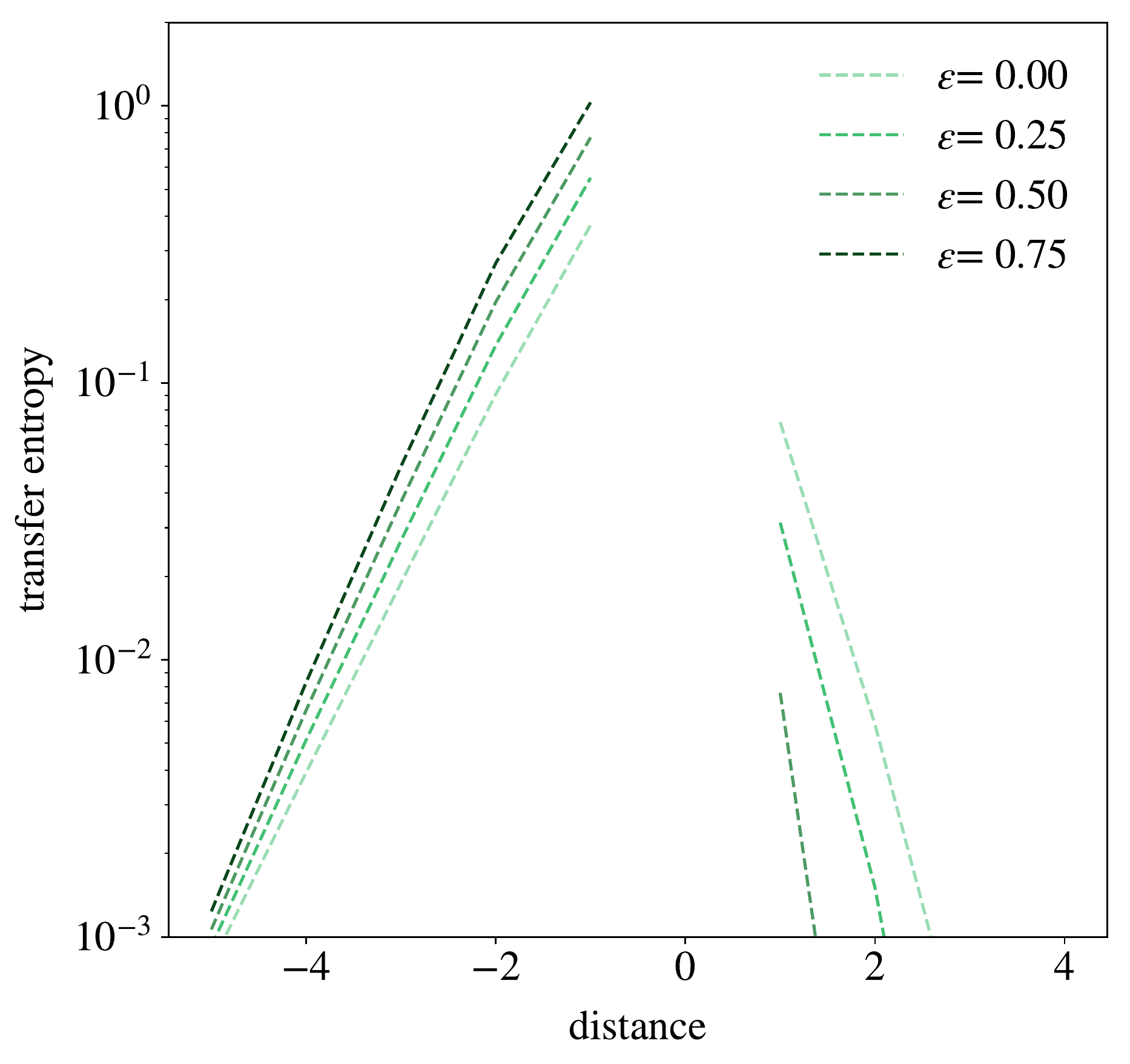}
 \caption{Transfer entropy in the linear case with noise gradient Eq.~\eqref{eq:linear_2}. 
 Here $\mathcal{T}_{i\to j}$ is plotted as a function of the oriented distance $j-i$, for different values of the parameter $\varepsilon$ ruling the asymmetry of the interactions.
 Parameters: $k=2$, $\gamma = 2.5$, $T=5$, temperature gradient $\Delta T = 1$,  $N=20$, $\Delta t = 10^{-2}$.\label{fig:linear_grad}}
\end{figure}

A slightly different situation is considered in Fig.~\ref{fig:linear_grad}. In this case the amplitudes of the noise terms are given by a piece-wise linear rule
\begin{equation}
\label{eq:linear_2} 
 T_i=T_0+|i|\Delta T
\end{equation} 
where $i$ ranges from $-N/2+1$ to $N/2$ and periodic boundary conditions are assumed. In this case we fix the variable $x_i$ with $i=N/4$ and repeat the previous analysis on RFs and TE. Not surprisingly, the behavior of $\matr{R}_{ij}$ (not shown) is identical to the one previously discussed for the case with homogeneous noise; this could be guessed by recalling that for linear models the response is only determined by the interaction matrix $\matr{A}$ and is independent of the choice of $\matr{B}$.
On the contrary, as we see in Fig.~\ref{fig:linear_grad}, the inhomogeneity in the noise terms makes the transfer entropies asymmetric even when the interactions between variables are completely symmetric ($\varepsilon=0$).
Again, this could be expected since correlations depend also on the noise matrix $\matr{B}$. Let us remark that this qualitative difference also occurs when using these observables as proxies of causation relations~\cite{baldovin2020understanding,auconi2021fluctuation,manshour21}.

\begin{figure}
 \centering
 \includegraphics[width=.8\linewidth]{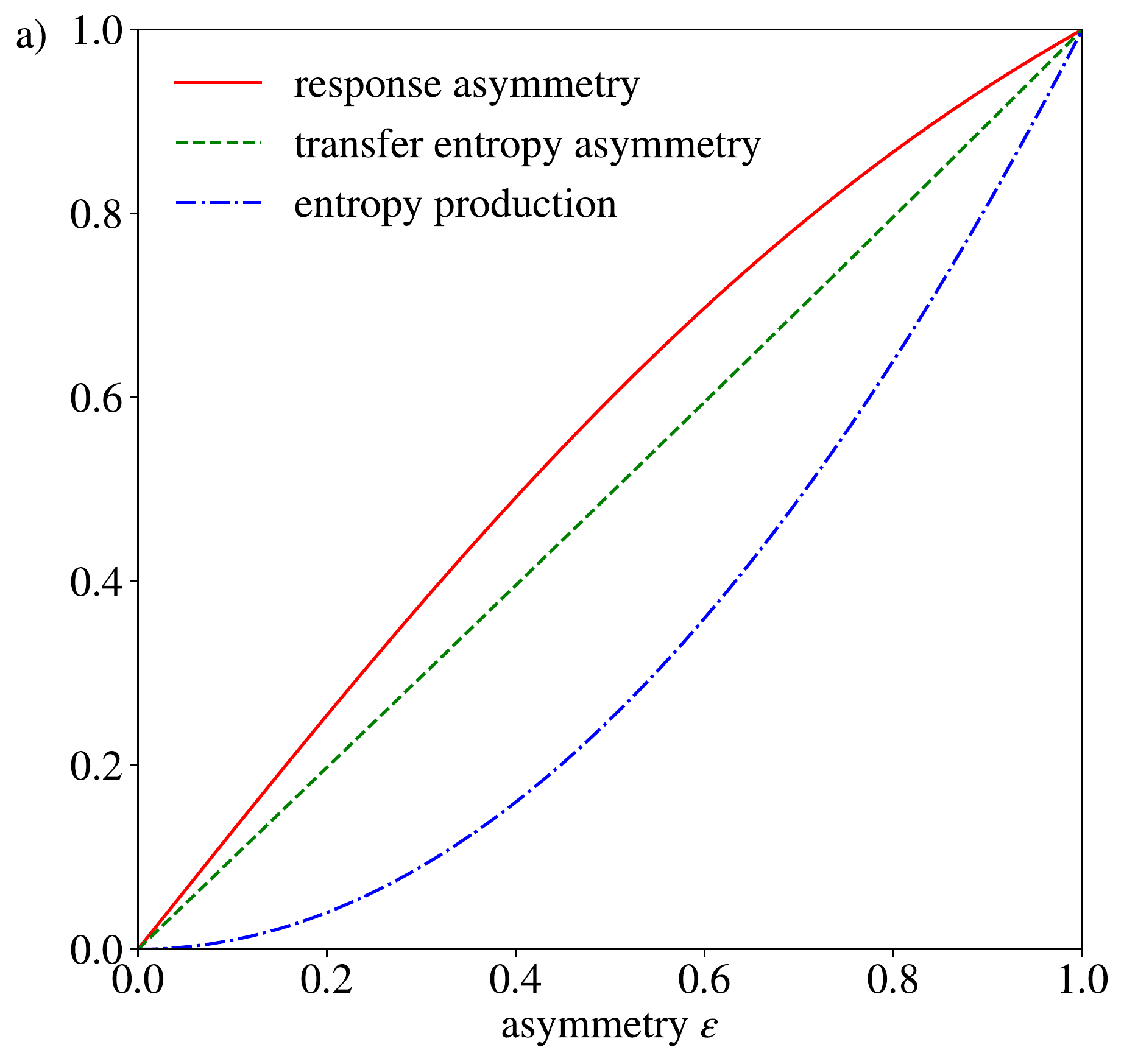}
 \includegraphics[width=.8\linewidth]{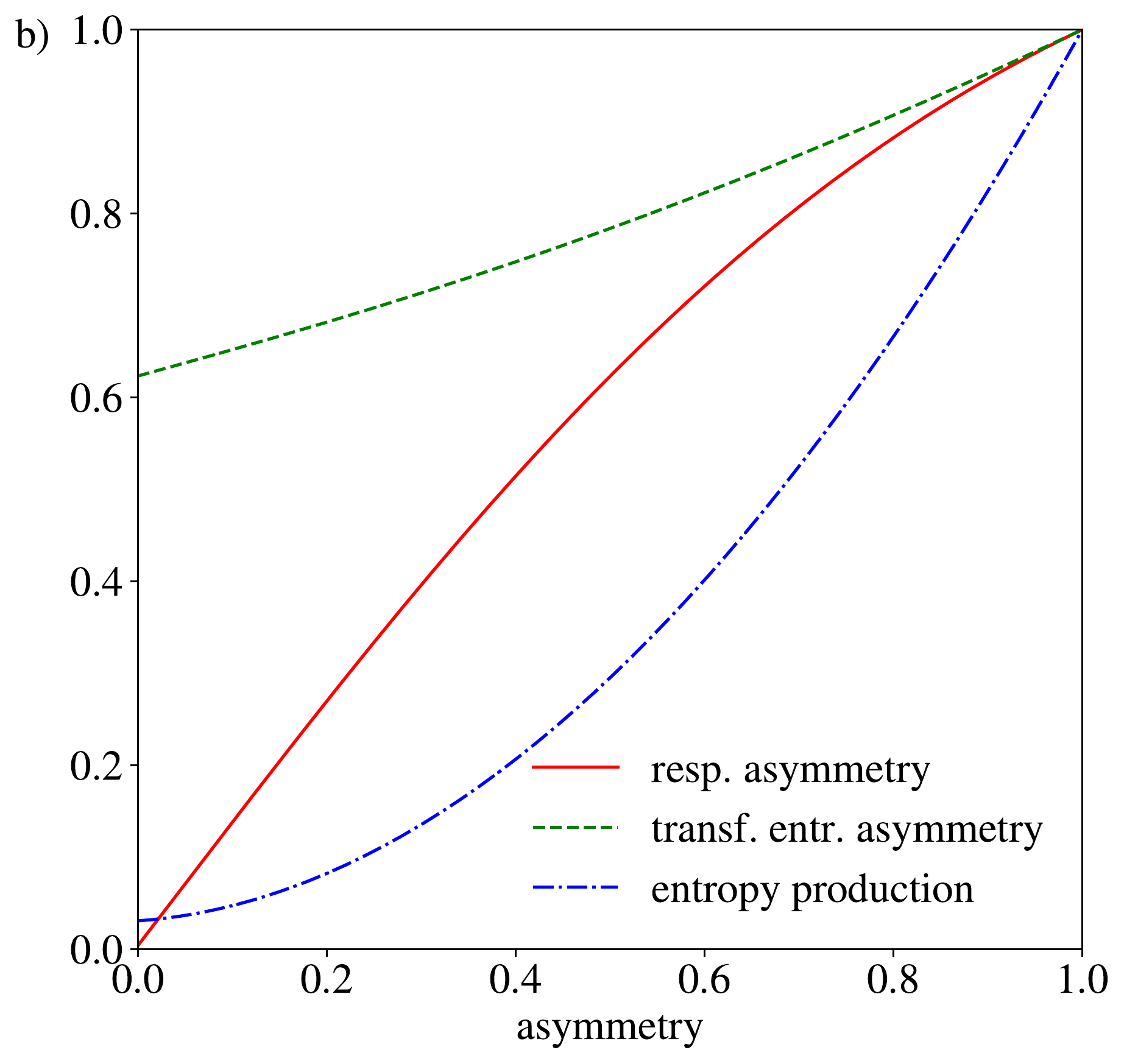}
 \caption{\label{fig:linear_comp}Comparison between entropy production rate and spatial asymmetry. 
 Panel (a): $\Sigma$, $A^{(\matr{R})}_i$ and $A^{(\mathcal{T})}_i$ for the system with uniform noise, as functions of the interaction asymmetry $\varepsilon$. All quantities are rescaled with their values at $\varepsilon=1$. Here the results are, by symmetry, independent of the choice of $i$. Panel (b): same for the system with noise gradient~\eqref{eq:linear_2}. Here $i=N/4$.
Parameters as in Figs.~\ref{fig:linear} and~\ref{fig:linear_grad}.}
\end{figure}

We now define the observables
\begin{equation}
 A^{(\matr{R})}_i=\sum_{j<i}\tilde{\matr{R}}_{ji}-\sum_{j>i}\tilde{\matr{R}}_{ji}
\end{equation}
and
\begin{equation}
 A^{(\mathcal{T})}_i=\sum_{j<i}\mathcal{T}_{i\to j}-\sum_{j>i}\mathcal{T}_{i\to j}\,,
\end{equation} 
which quantify the spatial asymmetry of the integrated response function and of the transfer entropy, respectively. In the homogeneous-noise case, summations have to be intended on $j=i-N/2,...,i-1$ and $i+1,...,i+N/2$ (where periodic boundary conditions are meant), and they, by symmetry, do not depend on $i$; in the case with a temperature gradient we will limit ourselves to $i=N/4$, and the summations are intended over $j=0, ..., N/4-1$ and $j= N/4+1,...,N/2$.

In Fig.~\ref{fig:linear_comp} such quantities are related to the lack of equilibrium measured by $\Sigma$, when varying $\varepsilon$. All observables are rescaled to their values at $\varepsilon=1$.
The meaning of Fig.~\ref{fig:linear_comp}(a) is quite clear: when the noise amplitudes are all equal, for $\varepsilon=0$ the system is at equilibrium and there is no breaking of spatial symmetry neither for the RFs nor for TE. 
This can be simply understood by recalling that the model can be seen as the overdamped dynamics of $N$ particles at temperature $T$ interacting via a quadratic potential, and the stationary state is represented by an equilibrium distribution.
As soon as $\varepsilon>0$, the dynamics has a ``preferred'' direction; as a consequence, detailed balance fails and $\Sigma$ grows. 
In this case, the failure of equilibrium is related to the breaking of spatial symmetries, revealed by the analysis of RF and TE. 

A different scenario emerges from the analysis of the model with different noise amplitudes Eq.~\eqref{eq:linear_2}. In this case even for $\varepsilon=0$ the system is out of equilibrium, as it is clear from the fact that $\Sigma>0$. 
Response functions and transfer entropy give qualitatively different information: the former are symmetric, as a consequence of the symmetry of interactions. The presence of probability currents is instead testified by the asymmetric behavior of the latter. We can conclude that, at least in linear systems, an asymmetry in the response functions can be seen as a genuine indicator of the (spatial) asymmetry of the dynamics, related to unbalanced internal interactions; TE, on the other hand, may show an asymmetric behavior even in presence of symmetric interactions, if the system is driven out of equilibrium by, e.g., a temperature gradient. Since in the following we will only consider systems with homogeneous noise, we expect the two quantities to provide similar information, and we will only discuss response functions.

\subsection{Effects of nonlinear terms}
\label{sec:nonlinear}

\begin{figure}
 \centering
 \includegraphics[width=.8\linewidth]{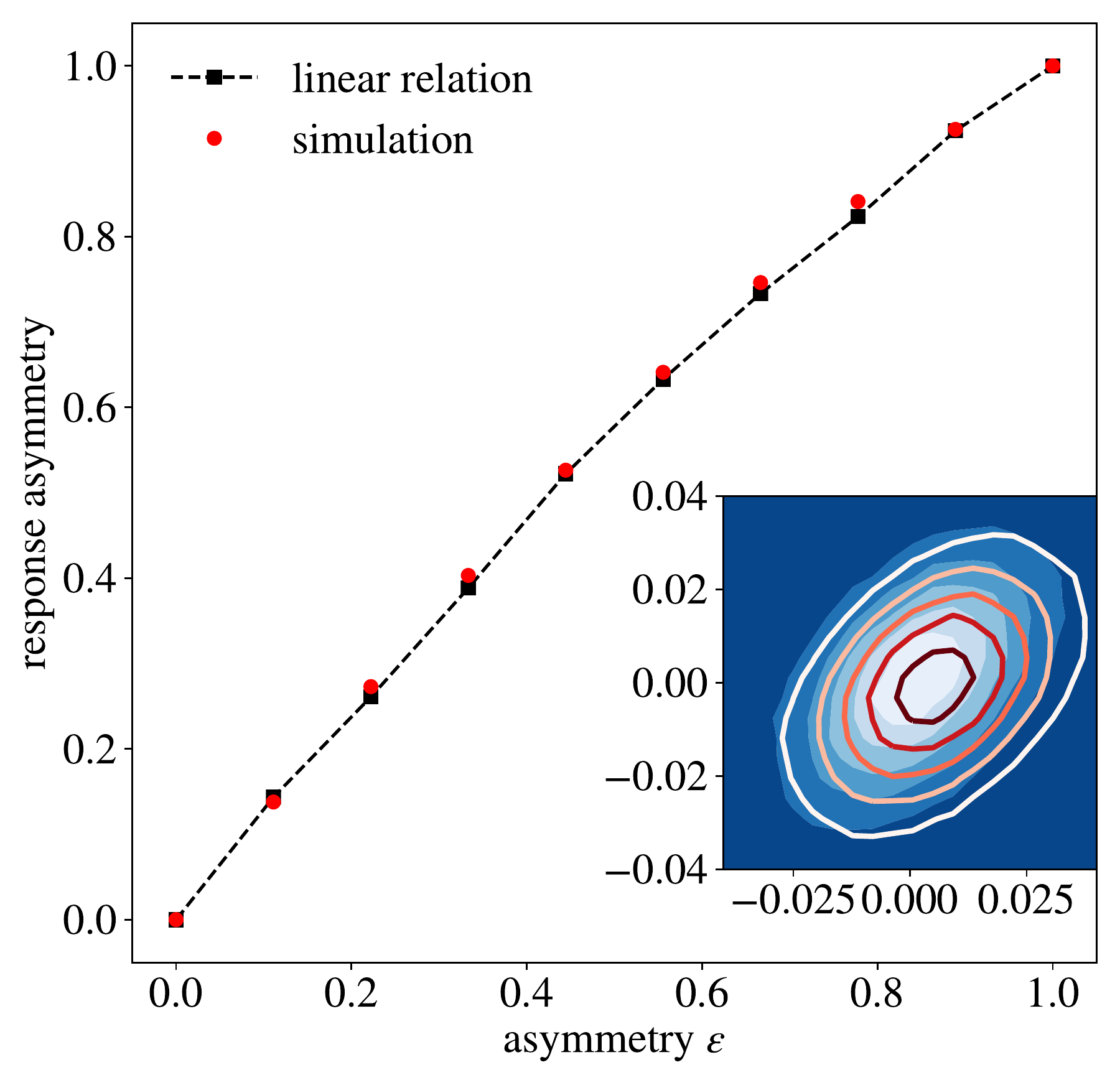}
 \caption{Ring model with non-linear terms.
  Main plot: asymmetry $A^{(\matr{R})}$ of the cumulative response as a function of the asymmetry in the interactions. Inset:
  two-variables pdf (lines are contour curves) and comparison with a Gaussian distribution with the same mean and covariance matrix (in blue). Parameters: $k=2$, $\gamma = 2.5$, $\zeta=10$, $T=0.5$,  $N=20$,
 For panel a), integration time $\Delta t =10^{-3}$, observation time $\tau = 50$ $\Delta t$, integration length $N_{\Delta T}=100$, repetitions $M=1000$, perturbation $\delta x=0.1$. For panel b): $\Delta t = 10^{-2}$, observation time $\tau = 8$ $\Delta t$, integration length $N_{\Delta T}=10$, repetitions $M=5 \cdot 10^4$.\label{fig:nonlinear}}
\end{figure}

To show that RFs are good indicators for the asymmetry of the dynamics, even in presence of nonlinear terms, in this paragraph we consider a particular case of the ring model described by Eq.~\eqref{eq:genmodel}, in which nonlinear terms explicitly appear in the drift functions~\eqref{eq:drift}:
\begin{equation}
 f(r)=\gamma r\,,\quad f_L(r)=\alpha r+\zeta r^3\,,\quad f_R(x)=\beta r+ \zeta r^3\,.
\end{equation}
Defining again $\varepsilon=(\beta-\alpha)/(2k)$, in Fig.~\ref{fig:nonlinear} we show how RFs depend on the asymmetry $\varepsilon$.

The results are quite similar to those shown in Fig.~\ref{fig:linear} for the linear case, suggesting (although, of course, not proving) that the presence of nonlinear terms does not hinder the ability of RFs to catch the privileged direction of the dynamics.

It is also worth noticing that, at least in this case, the RF can be fairly approximated with the linear combination of correlation functions suggested by Eq.~\eqref{intro-linear_regression}. 
This result is quite unexpected, since Eq.~\eqref{intro-linear_regression} only holds, in principle, for systems with linear interactions; when nonlinear drift terms are present one has to rely on the more general formula~\eqref{eq:fdr}, which may lead to nontrivial contributions. 
However, as discussed for instance in Ref.~\cite{baldovin2020understanding}, it should be noted that the r.h.s. of Eq.~\eqref{intro-linear_regression} is the linear regression between the state of the system at time $0$ and that at time $t$; it provides therefore the drift matrix $\matr{F}$ of the ``best'' linear approximation for the discrete-times dynamics
\begin{equation}
 \vect{x}_{s+t}=\vect{f}_t(\vect{x}_s)\simeq \matr{F}_t\vect{x}_s\,.
\end{equation} 
The RF associated with this approximated dynamics is exactly given by the matrix $\matr{F}_t$.

\section{Lorenz '96 model}
\label{sec:lorenz}

In 1996 E. Lorenz introduced a model to catch the essential features of atmospheric circulation, in the context of fluid dynamics~\cite{lorenz1996predictability}; it consists of a set of $N$ variables which evolve according to the following equations:
\begin{equation}\label{eq:c-lorenz_generale}
 \dot{x}_n = \mathcal{G}_n(\vect{x}) - \nu x_n +F\,,\quad
\quad n=1, \ldots, N\,,
\end{equation}
where 
\begin{equation}
 \mathcal{G}_n(\vect{x})=x_{n-1} \left(x_{n+1} - x_{n-2}\right)
\end{equation} 
and periodic boundary conditions $x_{n\pm N} = x_n$ hold.
The variables may represent a scalar meteorological quantity, such as vorticity or temperature, measured at equally spaced sites around a latitude circle \cite{karimi2010extensive}; the periodic boundary conditions are therefore natural, if we think of these variables as disposed on a circle along a parallel of the Earth.

The evolution of each variable is determined by three contributions \cite{lorenz1998optimal}: \textit{(i)} the nonlinear quadratic interaction term $\mathcal{G}_n(\vect{x})$, which is the only possible source of asymmetry of the dynamics; \textit{(ii)} a linear damping term, proportional to $-\nu$, that represents the mechanical or thermal dissipation and has the effect of reducing the total energy of the system; \textit{(iii)} a constant external force $F$ that prevents the total energy from decaying to zero and hence the stationary behavior from being trivial.
 In the case $\nu = 0$, $F=0$, which mimics the Euler's equation, there is a quantity conserved by the dynamics,
 \begin{equation}
 E = \sum_n^N x_n^2\,,
 \end{equation}
 and hence the states of the system lives on a ($N-1$)-dimensional surface. If instead $\nu \ne 0$ and $F \ne 0$, 
 one has a chaotic dissipative system. Asymptotically the motion evolves on a strange attractor with a fractal structure.
 For instance for $N=20$, $\nu = 1$, $F=5$ the Grassberger-Procaccia dimension is around 6.6~\cite{cecconi2012predicting}.
 For a detailed investigation of the system~\eqref{eq:c-lorenz_generale} in terms of linear response theory see~\cite{lucarini2018revising}.
 
Let us notice that the presence of chaos induces a subtle mathematical problem for the study of the generalized fluctuation-dissipation theorem, i.e. the fact that the invariant measure of a chaotic system can now be singular, against the hypotheses of the theorem (see Ref.~\cite{sarracino2019fluctuation}).
We can overcome this difficulty by adding some stochasticity to the system, in the form of a Gaussian white noise with zero mean:
\begin{equation}\label{eq:c-lorenz-turbo}
 \dot{x}_n= \mathcal{G}_n(\vect{x}) - \nu x_n + F + \sqrt{2 T}\eta_n\,,\quad n= 1, \ldots, N.
\end{equation} 
where $\langle \eta_n(t) \eta_m(s)\rangle=\delta_{nm}\delta(t-s)$.

\begin{figure}
 \centering
 \includegraphics[width=.8\linewidth]{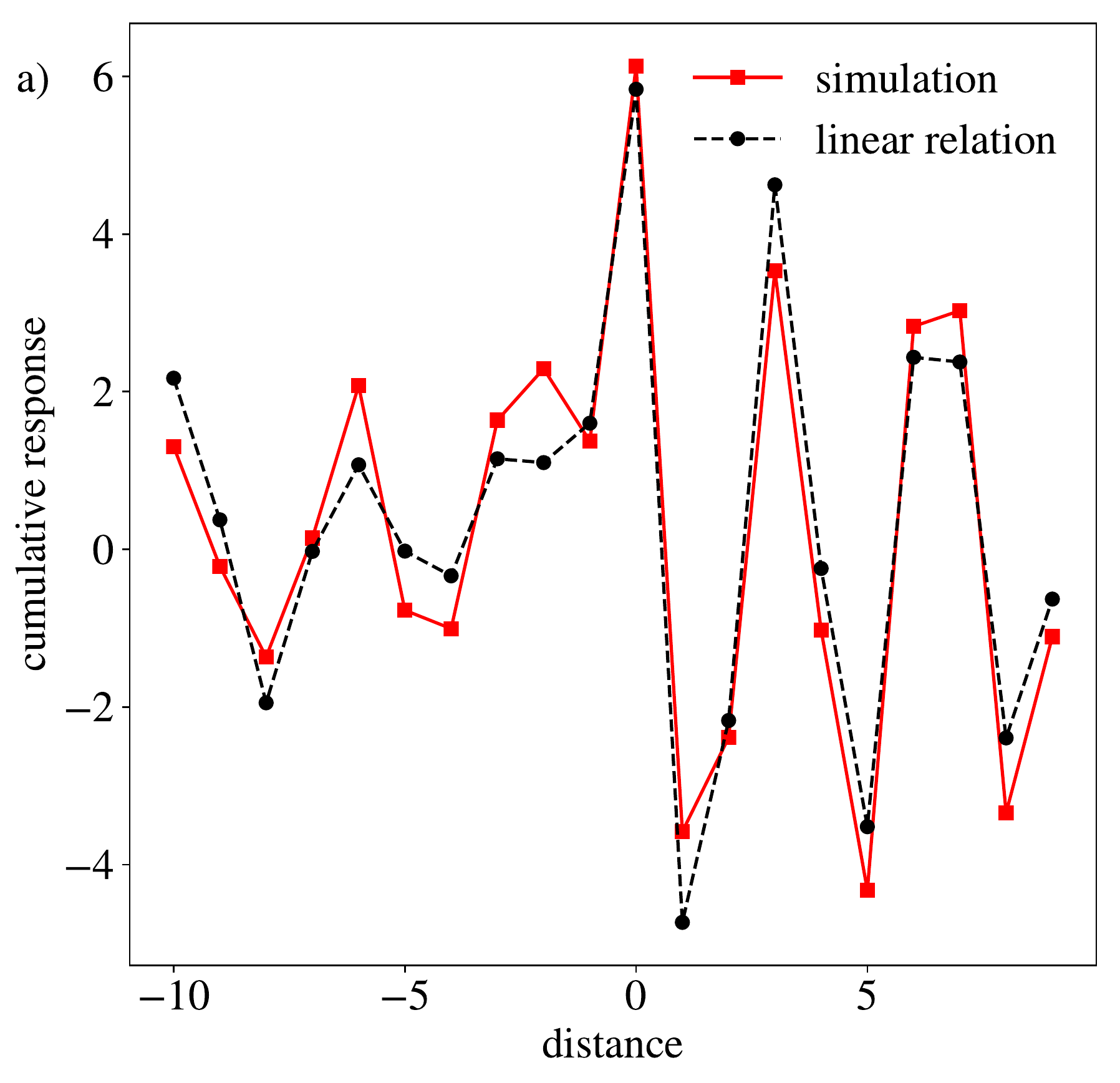}
\includegraphics[width=.8\linewidth]{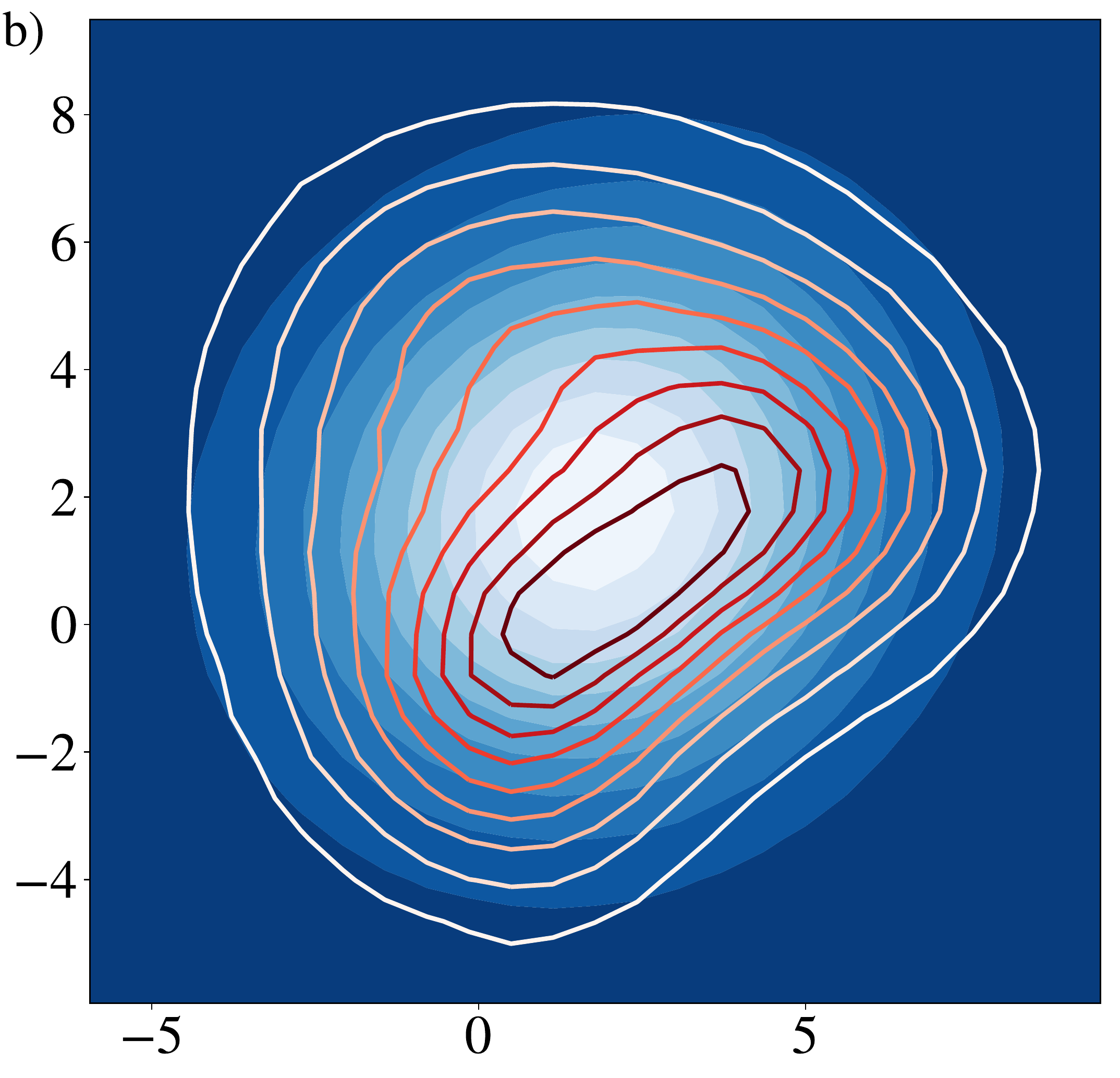}
 \caption{Lorenz '96 system in the turbulent regime. Panel (a): comparison between the actual cumulative responses of the system (squares) and the linear relation exactly  holding for Gaussian systems (circles), as in Eq. (\ref{lin-regr-cumu}). Lines are guides for the eye. Panel (b): joint pdf of two variables at distance $d=1$ (lines are contour curves) and the comparison with a Gaussian distribution with the same mean and covariance matrix (in blue). Parameters: $F=5$, $ N=20$, $\nu=1$, $T=2$, integration time $\Delta t=10^{-3}$, perturbation $\delta x=0.1$. \label{fig:lorenz_profile}}
\end{figure}

Figure~\ref{fig:lorenz_profile}(a) shows the numerical behavior of the cumulative correlation. The computation has been done by numerically integrating over a time interval long enough to allow correlations to decay to zero.
Indicating again by $\tilde{\matr{R}}$ and $\tilde{\matr{C}}$ the cumulative matrices obtained by integrating over time the elements of the matrices $\matr C$ and $\matr R$, respectively, we can test the validity of the linear approximation Eq.~\eqref{lin-regr-cumu}
that we know to hold exactly for linear systems. 
The behavior obtained from linear 
regression is qualitatively very similar to that obtained by integrating 
responses, even if, as in the case of ring models with non-linear interactions 
discussed before, the joint pdf of the considered variables is not Gaussian.
This suggests a handy way to measure spatial asymmetry even in systems with complex
dynamics, by only measuring suitable linear combinations of correlation functions.

The above results suggest that the Lorenz '96 dynamics is spatially asymmetric, namely the effects of the perturbation of a variable will have a preferential propagation direction.
It is interesting to study how this asymmetry depends on the model's parameters. 
One can introduce a constant $k$, inspired by Einstein's relation, such that
\begin{equation}\label{eq:k_def}
 \nu=k\nu_0\quad \quad T=kT_0\,.
\end{equation} 
It is worth noticing that in the limit $k \to \infty$ the nonlinear terms have a poor role, and this case corresponds therefore to a ``small Reynolds number'' fluid, whereas $k \to 0$ represents a ``large Reynolds number'' situation.

\begin{figure}
 \centering
 \includegraphics[width=.8\linewidth]{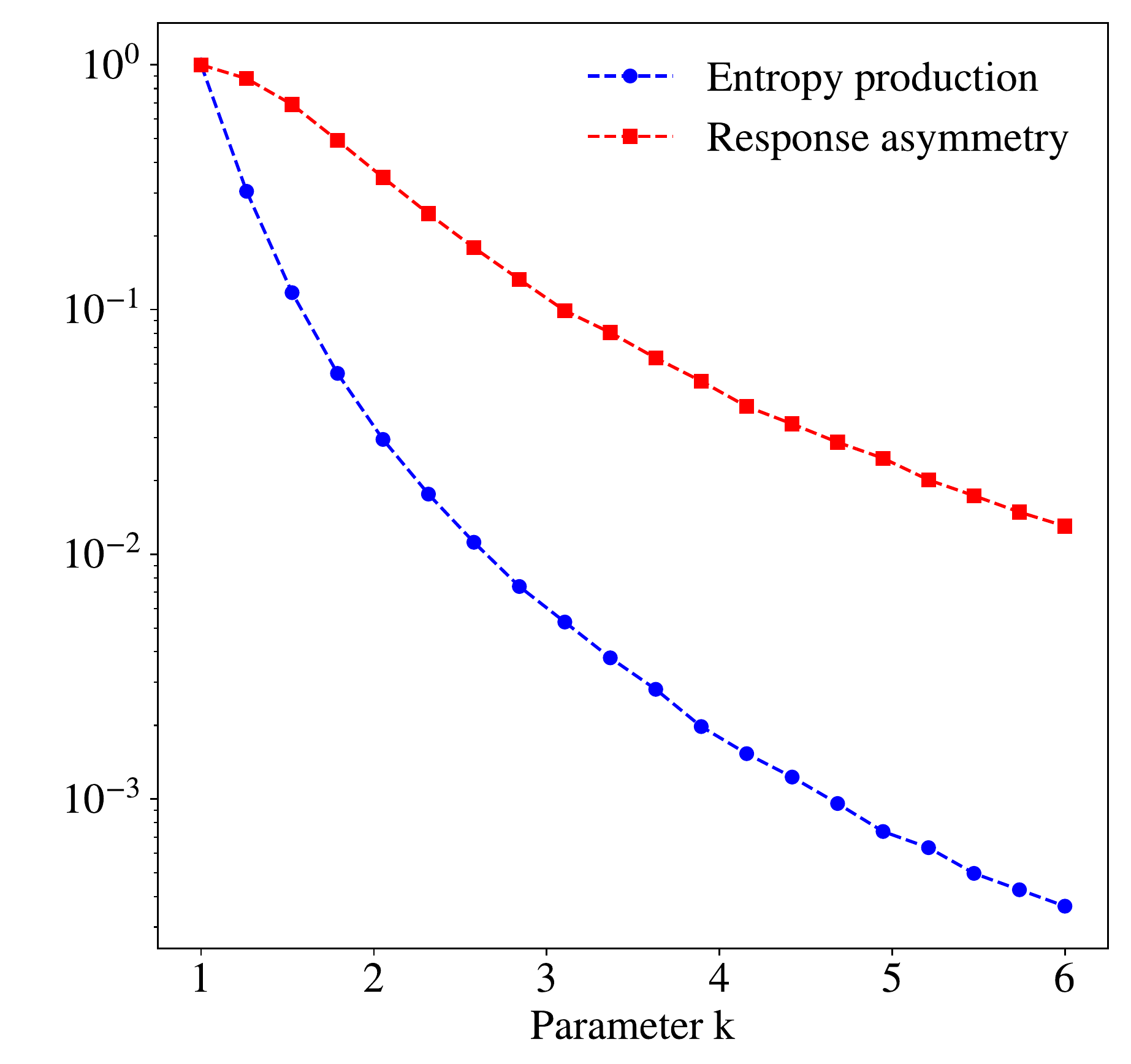}
 \caption{Entropy production and response asymmetry in Lorenz '96 model, as functions of parameter $k$ defined in Eq.~\eqref{eq:k_def}. Both quantities are rescaled with their values for $k=1$. Lines are guides for the eye.
 Parameters: $T_0 = \nu_0 = 2$, $F=5$, $N=20$, $\Delta t = 10^{-3}$, $M = 100$, $\delta x=0.1$.
 \label{fig:turbulent}}
\end{figure}

In Fig.~\ref{fig:turbulent} we show the behavior of the asymmetry of cumulative RFs and of the production of entropy $\Sigma$ as functions of  $k$.
We notice that in the large-$k$ limit, both quantities tend to zero, suggesting that the variables are becoming independent and the dynamics less asymmetric.

\section{Conclusions}
\label{sec:conclusions}

Whether and how the lack of equilibrium in a multi-dimensional system can be 
related to the presence of currents of measurable quantities, i.e. to a lack of 
spatial symmetry for the dynamics, is a wide and general problem of 
out-of-equilibrium statistical mechanics. Here we investigated the case of 
extended systems with one-dimensional periodic geometry, where spatial 
asymmetry can be simply characterized by measuring suitable dynamical 
observables such as RFs and TE, while the entropy production rate can be 
adopted, as usual, as a measure of the deviation from equilibrium.

We first tried to get some insight into their relation by looking at the 
behavior of linear ring models: in these systems nonequilibrium can be induced, 
for instance, by imposing asymmetric interaction terms or a ``temperature'' 
gradient, i.e. a spatial-dependent modulation of the amplitudes of the noise. In 
both cases we observe that the nonvanishing entropy production leads to an 
entropy current testified by the asymmetry of the TE; if the nonequilibrium 
is due to an asymmetry between interactions, also an analysis of RFs can
reveal it, and the qualitative scenario does not change by including nonlinear terms.

The same kind of analysis has been performed on more complex systems, leading to
similar conclusions. We considered the Lorenz '96 model in its ``turbulent'' version:
by varying a parameter which allows to switch from large to small Reynolds numbers,
we observe a transition from a out-of-equilibrium regime characterized by high asymmetry in the RFs
to a limit in which both time-reversal and spatial symmetries are restored.

Our results suggest the possibility to detect the absence of equilibrium in a generic system by studying
the asymmetry of suitable observables such as transfer entropy and, more importantly, response functions, which
are much simpler to measure; this possibility opens interesting perspectives and  deserves for sure further 
numerical and analytical investigation.

\acknowledgements{M.B. and A.V. acknowledge financial contribution of MIUR-PRIN2017 project \textit{Coarse-grained description for non-equilibrium systems and transport phenomena (CO-NEST)}. }

\appendix
\section{Transfer entropy for linear systems}
\label{sec:appendix_te}

Transfer entropy can be written explicitly only when the stationary distribution is known.
We are particularly interested in the Gaussian case, for which we report here the explicit results (see e.g.~\cite{sun2015causal} for a wider discussion).

Let us consider an $N$-variable system characterized by a multivariate Gaussian distribution, and let us denote
by $x$ and $y$ two of its variables. We know that
\begin{enumerate}
 \item The conditional probability of a Gaussian distribution is still a Gaussian with covariance matrix
 \begin{equation}\label{1}
 \Gamma_{x|y} = \Gamma_x - \Gamma_{x,y} \Gamma_y^{-1} \Gamma_{x,y}^T
 \end{equation}
 \item The entropy of an N-dimensional Gaussian distribution with covariance matrix $\Gamma$ is given by
 \begin{equation}\label{2}
 H = \frac{1}{2}\ln{|\Gamma|} + \frac{N}{2}\ln{2\pi e},
 \end{equation}
 where $|\Gamma|$ means the determinant of the matrix $\Gamma$.
\end{enumerate}
Starting from Eq. (\ref{def:transfer_entropy}), in the main text we are interested in the case in which $x_t \equiv x_i^{(t)}$ and  $y_t \equiv x_j^{(t)}$, so 
\begin{equation}
 \mathcal{T}_{j \to i}^{(t)}= H\left(x_i^{(t)}|x_i^{(t-1)}\right) - H\left(x_i^{(t)}|x_i^{(t-1)},x_j^{(t-1)}\right).
\end{equation}
Using Eq. (\ref{2}) we can write
   \begin{align}
 &\mathcal{T}_{j \to i}^{(t)}
= \frac{1}{2}\ln\left( \frac{\left|\Gamma_{x_i^{(t)}|x_i^{(t-1)}}\right|}{\left|\Gamma_{x_i^{(t)}|x_i^{(t-1)}, x_j^{(t-1)}}\right|}\right).
 \end{align} 
Given the stationarity of the system and using Eq. (\ref{1}), we can write the numerator and the denominator as combinations of correlations at the same time ($\sigma$) and after one time step ($C$), as
\begin{equation}
    \sigma_{x_i^{(t)}|x_i^{(t-1)}} = \sigma_{ii} - C_{ii} \sigma_{ii}^{-1} C_{ii}^{T}
\end{equation}
 and 
 \begin{equation}
       \sigma_{x_i^{(t)}|x_i^{(t-1)}, x_j^{(t-1)}} = \sigma_{ii} - \left(\begin{matrix} C_{ii} & C_{ij}\end{matrix} \right) \left(\begin{matrix} \sigma_{ii} & \sigma_{ij} \\\sigma_{ij} & \sigma_{jj} \end{matrix}\right)^{-1} \	\left( \begin{matrix} C_{ii} \\ C_{ij} \end{matrix}\right) 
 \end{equation}
 
We get
\begin{align}
\mathcal{T}_{j \to i}  &= \frac{1}{2} \ln{\left( \frac{\sigma_{ii} - \dfrac{ C_{ii}^2}{\sigma_{ii}}}{\sigma_{ii} - \left(\begin{matrix} C_{ii} & C_{ij}\end{matrix} \right) \left(\begin{matrix} \sigma_{ii} & \sigma_{ij} \\\sigma_{ij} & \sigma_{jj} \end{matrix}\right)^{-1} \	\left( \begin{matrix} C_{ii} \\ C_{ij} \end{matrix}\right) } \right)}=\nonumber\\
         &= \frac{1}{2} \ln{\left( \frac{\sigma_{ii} - \frac{ C_{ii}^2}{\sigma_{ii}}} {\sigma_{ii} - \frac{\sigma_{jj} C_{ii}^2-\sigma_{ij} C_{ij} C_{ii} -\sigma_{ij} C_{ii} C_{ij}+\sigma_{ii} C_{ij}^2}{\sigma_{ii}\sigma_{jj}-\sigma_{ij}^2} }\right)}=\nonumber\\
        &= \frac{1}{2}\ln{\left(1+\frac{\alpha_{ij}}{\beta_{ij} - \alpha_{ij}} \right)},
\end{align}
        
with
\begin{equation}
    \begin{cases}
        \alpha_{ij} = \left(\sigma_{ii} C_{ij}- C_{ii}\sigma_{ij}\right)^2\\
        \beta_{ij} = \left(\sigma_{ii}^2- C_{ii}^2\right)\left(\sigma_{ii}\sigma_{jj}-\sigma_{ij}^2\right)
    \end{cases}.
\end{equation}

Finally, if we want the expression for continuous-time systems, by following Eq. (\ref{continuous_time_TE}), we have
\begin{equation}\label{eq. Transfer Entropy}
 \mathcal{T}_{j \to i} = \lim_{t \to 0} \frac{1}{2 t} \ln{\left(1+ \frac{\alpha_{ij}(t)}{\beta_{ij}(t)-\alpha_{ij}(t)}\right)},
\end{equation}
with
\begin{equation*}
 \begin{cases}
 \alpha_{ij}(t) = \left( \sigma_{ii} \matr{C}_{ij}(t) -\sigma_{ij} \matr{C}_{ii}(t) \right)^2\\
 \beta_{ij}(t) = \left(\sigma_{ii}^2 - \matr{C}_{ii}(t)^2 \right) \left(\sigma_{ii} \sigma_{jj} - \sigma_{ij}^2\right)\,,
 \end{cases}
\end{equation*}
which is the expression we used in our numerical study.

\bibliography{biblio}
\end{document}